\documentclass[11pt, a4paper]{article}
\pdfoutput=1

\usepackage{jcappub}

\usepackage{comment}
\usepackage{amsmath}
\usepackage{amsfonts}
\usepackage{amssymb}
\usepackage{graphicx}
\usepackage{mathrsfs}
\usepackage{latexsym}
\usepackage{color}
\usepackage{slashed, cancel}
\usepackage{hyperref}
\usepackage{cleveref}
\usepackage{float}
\usepackage{tikz}
\usepackage{bbold}
\usepackage{soul} 
\usepackage{mathtools}
\usepackage{support-caption} 
\usepackage{subcaption}
\usepackage{enumitem}
\usepackage{journals}
\allowdisplaybreaks

\hypersetup{urlcolor=blue, colorlinks=true} 

\usepackage{fancyhdr}
\numberwithin{equation}{section}

\definecolor{rossos}{rgb}{0.8,0.2,0.3}
\definecolor{bluscuro}{rgb}{0.15, 0.2, .85}
\definecolor{bluchiaro}{cmyk}{1,.3,0.,0.1}
\hypersetup{colorlinks, citecolor=bluscuro, linkcolor=bluscuro, urlcolor=bluscuro}

\makeatother   
\newcommand{\GeV}{{\rm \,GeV}}
\newcommand{\TeV}{{\rm \,TeV}}

\newcommand{\cm}{{\rm \,cm}}

\newcommand{\mathsc}[1]{\text{\textsc{#1}}}

 \def\be   {\begin{equation}}   \def\ee   {\end{equation}}
 \def\ba   {\begin{array}}      \def\ea   {\end{array}}
 \def\bea  {\begin{eqnarray}}   \def\eea  {\end{eqnarray}}
 \def\bean {\begin{eqnarray*}}  \def\eean {\end{eqnarray*}}

\begin{document}

\today

\title{Loop Effects in Direct Detection}

\author{Nicole F.\ Bell,}
\author{Giorgio Busoni and}
\author{Isaac W. Sanderson}
\affiliation{ARC Centre of Excellence for Particle Physics at the Terascale \\
School of Physics, The University of Melbourne, Victoria 3010, Australia}

\emailAdd{\tt n.bell@unimelb.edu.au}
\emailAdd{\tt giorgio.busoni@unimelb.edu.au}
\emailAdd{\tt isanderson@student.unimelb.edu.au}

\abstract{ 
We consider loop level contributions to dark matter scattering off nucleons in cases where the spin independent scattering cross section is absent or suppressed at tree level. In the case of a pseudoscalar interaction, for which the tree level cross section is both spin-dependent and suppressed by 4 powers of the exchanged momentum, we show that loop diagrams give rise to a non-zero spin independent cross section. Importantly, if the pseudoscalar interaction is formulated using a gauge invariant framework, loop effects generate an effective $\overline{\chi}\chi h$ vertex and result in a scattering cross section that is within reach of current or forthcoming experiments.
We also consider the case of inelastic dark matter, for which the tree-level direct detection cross section is negligible when the inelastic $\chi_1 N \rightarrow \chi_2 N$ process is kinematically suppressed. In this case, loop diagrams generate an interaction with both initial and final $\chi_1$ states and hence permit measurable, spin independent, $\chi_1 N \rightarrow \chi_1 N$ elastic scattering.
As such, we are able to probe parameter space that was previously considered inaccessible to direct detection.}
\maketitle


\section{Introduction}

While the cosmological evidence for Dark Matter (DM) is mounting, particle physics experiments are yet to observe a signal that can be conclusively attributed to a DM particle. Among the plethora of DM particle candidates exists the well-motivated Weakly Interacting Massive Particle (WIMP), which has been a staple in collider, direct and indirect searches for DM.

Direct Detection (DD) experiments attempt to measure nuclear recoils from the scattering of galactic DM off terrestrial matter. Given the limited speed of DM, $v \sim 10^{-3} c$, momentum transfers in such scattering processes are no more than a few hundred MeV -- far below the energy scale at which the new physics is expected to emerge in the typical WIMP paradigm. For this reason, DM interactions are efficiently modelled via an Effective Field Theory (EFT) approach \citep{Fitzpatrick:2012ix,Fitzpatrick:2012ib,Anand:2013yka}, in which higher energy degrees of freedom are integrated out to give a series of higher dimensional, non-renormalisable Lorentz structures. 

Direct Detection constraints are the strongest for unsuppressed Spin Independent (SI) scattering. Unfortunately, the SI scattering cross section is absent or suppressed in many theories. For example, those with tree level axial vector or pseudoscalar interactions have velocity/momentum suppressed or vanishing SI scattering, respectively. While axial vector interactions allow unsuppressed Spin Dependent (SD) scattering at tree level, the existing constraints on the cross section are several orders of magnitude weaker than for a SI interaction. Pseudoscalar interactions not only lack a tree level SI scattering interaction, but even the SD scattering is momentum suppressed, making them notorious for evading DD constraints.

While EFT descriptions of these interactions are appropriate for DD, such a framework potentially loses validity at collider energies \citep{Busoni:2013lha, Busoni:2014sya, Busoni:2014haa,
  Buchmueller:2013dya, Shoemaker:2011vi,Bell:2016obu,Abdallah:2014hon}. In this context, they were thus replaced in favour of Simplified Models \citep{Abdallah:2014hon,Abdallah:2015ter,Abercrombie:2015wmb,Boveia:2016mrp,Jacques:2016dqz,Buckley:2014fba,Harris:2014hga} which, in addition to the DM candidate, usually contain a single spin-0 or spin-1 particle that mediates the interaction between visible matter and DM.
Such single mediator Simplified Models, however, can suffer from issues related to breaking gauge invariance, unitarity, and renormalisability \citep{Bell:2015sza, Bell:2015rdw, Haisch:2016usn,
  Englert:2016joy, Kahlhoefer:2015bea, Bell:2016fqf, Ko:2016zxg,
  Duerr:2016tmh, Bell:2016uhg}. For example, a single neutral spin-0 mediator cannot couple to both SM fermions and fermionic DM -- doing so would require it both to be charged under SU(2)$_L$, to couple to SM particles, and to be a SM singlet, in order to couple to DM -- a problematic contradiction.

An appealing means to model the spin-0 mediator interaction in a gauge invariant way is to introduce both a spin-0 singlet (either an CP-even scalar or CP-odd pseudoscalar) as well as an additional Higgs doublet.  After SU(2) symmetry breaking, the (pseudo)scalar singlet mixes with the (pseudo)scalar component of the doublet, resulting in two mixed (pseudo)scalar mediators which couple the visible and dark sectors~\citep{Ipek:2014gua,Berlin:2015wwa,Goncalves:2016iyg,No:2015xqa,Haisch:2016gry,Bauer:2017ota,Tunney:2017yfp, Baek:2017vzd}. 

Direct detection constraints on the scalar and pseudoscalar mediator models are very different. The tree level exchange of a CP-even scalar results in a large SI scattering cross section, and hence stringent constraints (though the gauge invariant model allows interesting interference effects)~\citep{Bell:2016ekl, Bell:2017rgi}. In the case of a CP-odd scalar, however, the tree-level scattering cross section is negligible~\citep{Arcadi:2017wqi}, as it is a momentum suppressed, spin-dependent process.  However, although the SI cross section is absent at tree level, a non-zero contribution can be induced at loop level. For the gauge invariant implementation of the pseudoscalar model, an an effective $h\chi\chi$ interaction is induced at loop level~\citep{Ipek:2014gua}, which permits a measurable scattering of DM from nucleons.

Inelastic DM is another model in which DD is suppressed at tree level~\citep{Chang:2008gd, Cui:2009xq, ArkaniHamed:2008qn}. These models feature a pair of DM states, $\chi_1$ and $\chi_2$, with a tiny mass splitting, $\delta m = m_{\chi_2} - m_{\chi_1} \ll m_{\chi_i}$.  Because the couplings have an off-diagonal ($\chi_1$-$\chi_2$) structure, tree level DD can occur only by scattering the lighter $\chi_1$ to the heavier $\chi_2$ through the inelastic $\chi_1 N \rightarrow \chi_2 N$ process. This results in strong kinematic suppression of the cross section. Given the low velocity of galactic DM, $\delta m$ is required to be $\lesssim \mathcal{O} (10-100 \text{ keV})$ for this process to be kinematically allowed in conventional DD experiments. Inelastic scattering was used to offer a possible explanation for the tension between DAMA and other DD experiments~\citep{DelNobile:2015lxa,Scopel:2015eoh}, as the inelastic suppression differs for each experiment due the different target nucleon masses~\citep{TuckerSmith:2001hy}. Inelastic DM can arise in various existing models, e.g.~Sneutrino CDM in supersymmetric theories containing lepton number violation can produce the correct relic density, despite having only off-diagonal couplings for scattering \citep{Hall:1997ah}. More generally, it arises naturally in models where the DM is pseudo-Dirac. We shall see that although the inelastic DD process is kinematically suppressed, loop effects introduce an 
unsuppressed elastic scattering interaction.

This paper will demonstrate that, because the tree level DD process is highly suppressed, loop level diagrams will dominate the scattering cross section for both the gauge invariant pseudoscalar Simplified Model and inelastic DM.  Importantly, these loop diagrams generate unsuppressed SI scattering, at a level that may be observable in forthcoming experiments.
Furthermore, not only does the loop diagram for the gauge invariant pseudoscalar model remove the significant momentum suppression of the cross-section, it also removes any dependence on the Yukawa couplings of the $SU(2)$ pseudoscalar to SM fermions. Given that the DM-pseudoscalar coupling has to be large enough for thermal freezeout to produce the correct abundance, this offers the opportunity for non-negligible constraints for DD with pseudoscalar mediators.

We outline the gauge invariant pseudoscalar mediator and inelastic DM models in Section~\ref{sec:model}, and present our results for the loop level scattering operators.  We derive current DD constraints and future projections in Section~\ref{sec:constraints} and present
our conclusions in Section~\ref{sec:conclusions}.

\section{Example Models}
\label{sec:model}
\subsection{Gauge-Invariant Pseudoscalar Mediator}
\label{sec:modelP}
The model we outline below was first presented in \citep{Ipek:2014gua}, and has been discussed more recently in \citep{Baek:2017vzd,Bauer:2017ota,Tunney:2017yfp,Arcadi:2017wqi} as the simplest gauge-invariant formulation of a model with an $s$-channel pseudoscalar mediator portal to DM. A scalar version of the model was instead discussed in \citep{Bell:2016ekl,Bell:2017rgi}. The model has an additional Higgs doublet and an additional SM singlet pseudoscalar particle, $P$, coupling to DM with pseudoscalar couplings. The singlet pseudoscalar mixes with the pseudoscalar contained in the additional Higgs doublet through a cubic interaction term in the potential. In this way the model has two pseudoscalar mediators connecting the visible and dark sectors. The potential reads
\begin{equation}
V(\Phi_h,\Phi_H,P) = \hat{V}_{\mathsc{2hdm}}(\Phi_h,\Phi_H) + V_P(P) + V_{P\mathsc{2hdm}}(\Phi_h,\Phi_H,P), \label{eq:potential}
\end{equation}
where\nobreak
\bea
\hat{V}_{\mathsc{2hdm}}(\Phi_h,\Phi_H) &=& \hat{M}_{11}^2 \Phi_h^\dagger \Phi_h + \hat{M}_{22}^2 \Phi_H^\dagger \Phi_H +  (\hat{M}_{12}^2 \Phi_H^\dagger \Phi_h + h.c.) + \frac{\hat{\lambda}_1}{2}(\Phi_h^\dagger \Phi_h)^2 + \frac{\hat{\lambda}_2}{2} (\Phi_H^\dagger \Phi_H)^2 \nonumber \\
&+&\hat{\lambda}_3 (\Phi_h^\dagger \Phi_h)(\Phi_H^\dagger \Phi_H) + \hat{\lambda}_4 (\Phi_H^\dagger \Phi_h)(\Phi_h^\dagger \Phi_H)  
+ \frac{\hat{\lambda}_5}{2} \left( (\Phi_H^\dagger \Phi_h)^2 + h.c.\right),\\
V_P(P) &=& \frac{1}{2} M_{PP}^2 P^2 + \frac{1}{4} \lambda_P P^4, 
\\
V_{P\mathsc{2hdm}}(\Phi_h,\Phi_H,P) &=& 
(i\mu_{hHP} \Phi_h^\dagger \Phi_H P + h.c.).
\label{eq:s2hdm}
\eea
The 2HDM potential has been denoted as $\hat{V}_{\mathsc{2hdm}}$, and expressed in terms of $\hat{\lambda}_i$, to make manifest that we are writing it in the Higgs basis, where $\langle\Phi_H\rangle =0$\footnote{This choice is different from the one of \citep{Baek:2017vzd,Bauer:2017ota,Tunney:2017yfp}, where the potential is instead expressed in terms of the fields $\Phi_{1,2}$, which are rotated by an angle $\beta$ with respect to $\Phi_{h,H}$. The $\mathcal{Z}_2$ symmetry of the Yukawa terms is manifest in the $\Phi_{1,2}$ basis. The relationship between the couplings in the $\Phi_{1,2}$ basis, $\hat{\lambda}_i$, and those in the $\Phi_{h,H}$ basis, $\lambda_i$, can be found in~\citep{Bell:2017rgi}.}. In the alignment limit\footnote{The alignment limit can be obtained either by tuning the parameters of the model depending on the value of $\tan\beta$, or by imposing some symmetry. For example, with a CP2 symmetry\citep{Dev:2014yca} the alignment limit is naturally obtained for any value of $\tan\beta$, and the 2HDM potential also becomes invariant under rotations of the 2 doublets.}, the fields can be expressed in terms of mass eigenstates in the following way:
\bea
\Phi_h &=& \cos\beta \Phi_1 + \sin\beta \Phi_2 = \left(
\begin{array}{cc}
 G^+ \\
\frac{v + h + i G^0}{\sqrt{2}} \\
\end{array}
\right),\label{eq:alignh}\\
\Phi_H &=& -\sin\beta \Phi_1 + \cos\beta \Phi_2 = \left(
\begin{array}{cc}
 H^+ \\
\frac{ H + i R}{\sqrt{2}} \\
\end{array}
\right),\\
R &=& \cos\theta A - \sin\theta a,\\
P &=& \sin\theta A + \cos\theta a,
\eea
where the mixing angle depends on the value of the cubic coupling $\mu_{hHP}$ via the following relation:
\begin{equation}
\sin2\theta = \frac{\mu_{hHP} v}{M_{A}^2-M_{a}^2}.
\end{equation}
For more details about the model, see \citep{Ipek:2014gua}.

%
%
%

\begin{figure}[t]
    \centering
    \includegraphics{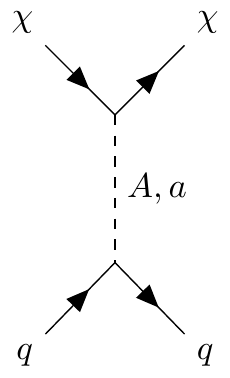} 
    \caption{Spin-dependent DM-nucleon scattering arises from the tree level exchange of the pseudoscalar mediators.}
    \label{fig:feynDDtreePS}
\end{figure}

At tree level, the model generates a very small spin-dependent cross section, suppressed by $q_{tr}^4$. Tree level scattering takes place via the exchange of $A$ or $a$, as shown in Fig.~\ref{fig:feynDDtreePS}.  In the EFT limit, the relevant scattering operator is 
\begin{equation}
O_4^N = \bar{\chi}\gamma^5 \chi \bar{N}\gamma^5 N,  \label{eq:op4}
\end{equation}
with a coefficient of
\begin{equation}
c_N^4 = m_N\frac{y_\chi \cos\theta\sin\theta}{v}\left(\frac{1}{M_{S_1}^2}-\frac{1}{M_{S_2}^2}\right)\sum_{q=u,d,s} \left(\epsilon_q-\left(\epsilon_u+2\epsilon_d\right)\frac{\bar{m}}{m_q}\right)\Delta_q^{(N)}.\label{eq:c4}
\end{equation}
where $\bar{m}^{-1}=\frac{1}{m_u}+\frac{1}{m_d}+\frac{1}{m_s}$, while $\Delta_q$ are  are defined, for example, in \citep{DelNobile:2013sia}.
DD constraints on the operator \ref{eq:op4} with coefficient \ref{eq:c4} are exceedingly weak, and translate to lower bounds on $M_a$ that are smaller than $1\GeV$ \citep{DelNobile:2013sia}.

\begin{figure}
    \centering
    \includegraphics[width=0.3\textwidth]{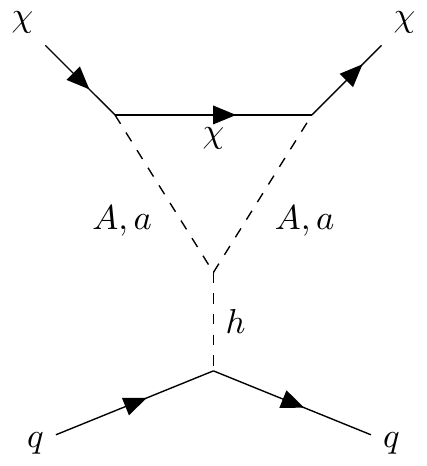} \hspace{0.02\textwidth}
    \includegraphics[width=0.3\textwidth]{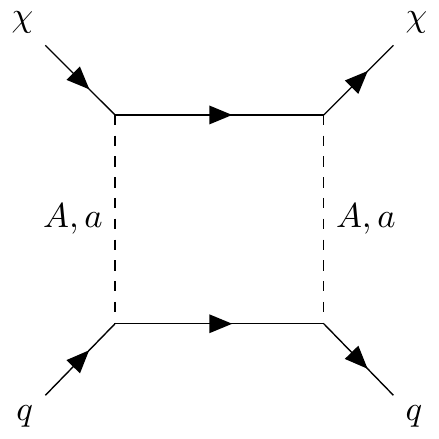} \hspace{0.02\textwidth}
    \includegraphics[width=0.3\textwidth]{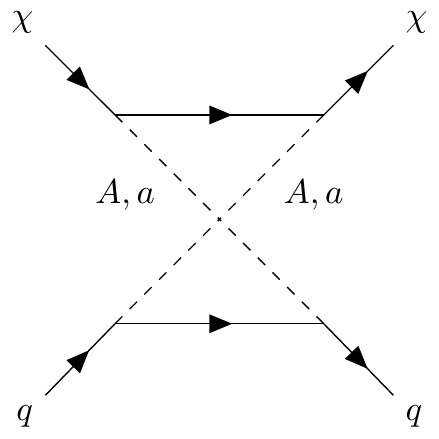} 
    \caption{Spin-independent DM-nucleon scattering arises from the loop exchange of the mixed pseudoscalar mediators. Left panel: triangle diagrams. Central and right panel: box diagrams.}
    \label{fig:feynDDPS}
\end{figure}

At loop level, however, an unsuppressed spin-independent cross section is generated by the diagrams in Fig.~\ref{fig:feynDDPS}. The triangle diagrams in the left panel of Fig.~\ref{fig:feynDDPS} are proportional to $m_q$ while the box diagrams in the central and right panels of Fig.~\ref{fig:feynDDPS} are proportional to $m_q^3$, thus the box diagrams are sub-leading as found in \citep{Ipek:2014gua} (unless Type II with $\tan\beta\gtrsim50$).
The triangle diagram does not depend on the Yukawa sector of the 2HDM.

The low energy effective operator in the approximation $M_{A}\gg M_{a}$ (in which case one considers only the diagram with two $a$'s appearing in the loop) is
\begin{eqnarray}
\mathcal{L} &=&  -\frac{y_\chi^2 m_q m_\chi \cos^2\theta \left(\left(M_{A}^2-M_{a}^2\right)\sin^2 2\theta - 2\hat{\lambda}_{34-5} v^2 \sin^2\theta\right)}{32\pi^2 m_h^2 m_\chi^2 v^2}     F_1\left(\frac{m_\chi^2}{M_{a}^2}\right) \bar{\chi}\chi\bar{q}q, \label{eq:looppsapprox}\\
F_1(x) &=& \int_0^1 dz \frac{x (1-z) z}{x z^2-z+1} = \frac{(6 x-2) \log \left(\frac{\sqrt{1-4 x}+1}{2 \sqrt{x}}\right)+\sqrt{1-4 x} ((x-1) \log (x)-2 x)}{2 \sqrt{1-4 x} x}, \label{eq:f1}
\end{eqnarray}
Our result differs from the one found in \citep{Ipek:2014gua} because of two reasons. First, we find that the amplitude contains an additional $\cos^2\theta$ factor, coming from the $\bar{\chi}\chi a$ Yukawa couplings, on the top of the $\sin^2 2\theta$ factor coming from the cubic scalar vertex $aah$. Second, we find that the cubic scalar vertex $aah$ receives contributions not only from the portal term $\frac{1}{2}(i\mu_{hHP} \Phi_1^\dagger \Phi_2 P + h.c.)$, but also from the 2HDM terms proportional to $\hat{\lambda}_{3},\hat{\lambda}_{4}$, and $\hat{\lambda}_{5}$, as it is manifest in Eq. \ref{eq:looppsapprox}. Depending on the point of the parameter space, these differences can be significant. To reduce the number of free parameters, we will fix $\lambda_{34-5}=\hat{\lambda}_3+\hat{\lambda}_4-\hat{\lambda}_5=\frac{m_h^2}{v^2}$ through all the paper\footnote{Note that this relation is obtained in the case of the CP2 symmetry considered in \citep{Dev:2014yca}, by also imposing an additional constraint $\hat{\lambda}_5=0$.}, as a benchmark point.

When considering all diagrams, one instead obtains  \\ 
\begin{eqnarray}
\mathcal{L}_{eff} &=& - \frac{y_\chi^2 m_q m_\chi}{16\pi^2 m_h^2 v^2}     G\left(\frac{m_\chi^2}{M_{A}^2},\frac{m_\chi^2}{M_{a}^2},\frac{m_h^2}{m_\chi^2},\theta\right) \bar{\chi}\chi\bar{q}q,\label{eq:loopps}\\
G\left(x,y,z,\theta\right) &=& F_1(x)\sin^2\theta\hat{\mu}_{AAh}+F_1(y)\cos^2\theta\hat{\mu}_{aah}+F_2(x,y)\sin2\theta\hat{\mu}_{Aah},\label{eq:f2}\\
\mu_{AAh} &=& \frac{1}{2}\sin^2(2\theta)\left(\frac{1}{x}-\frac{1}{y}\right) +z\frac{\hat{\lambda}_{34-5}v^2}{m_h^2}\cos^2\theta \rightarrow -\frac{1}{2}\sin^2(2\theta)\left(\frac{1}{x}-\frac{1}{y}\right) +z\cos^2\theta, \label{eq:mu11h}\\
\mu_{Aah} &=& \frac{1}{4}\sin(4\theta)\left(\frac{1}{x}-\frac{1}{y}\right) -\frac{z}{2}\frac{\hat{\lambda}_{34-5} v^2}{m_h^2}\sin(2\theta) \rightarrow -\frac{1}{4}\sin(4\theta)\left(\frac{1}{x}-\frac{1}{y}\right) -\frac{z}{2}\sin(2\theta), \quad  \quad  \quad  \label{eq:mu12h}\\
\mu_{aah} &=& - \frac{1}{2}\sin^2(2\theta)\left(\frac{1}{x}-\frac{1}{y}\right) +z\frac{\hat{\lambda}_{34-5}v^2}{m_h^2}\sin^2\theta \rightarrow \frac{1}{2}\sin^2(2\theta)\left(\frac{1}{x}-\frac{1}{y}\right) +z\sin^2\theta,\label{eq:mu22h}\\
F_2\left(x,y\right) &=& \int_0^1 dz \frac{x y z \log \left(\frac{x y z^2-y z+y}{x y z^2-x z+x}\right)}{y-x} \nonumber\\
= \frac{1}{4xy(x-y)} &\bigg(&x^2 ((2 y-1) \log (y)-2 y)+x^2 \sqrt{1-4 y} \left(\log (4 y)-2 \log \left(\sqrt{1-4 y}+1\right)\right) \nonumber\\
&-&2 x y^2 (\log (x)-1)+y^2 \log (x)+\sqrt{1-4 x} y^2\left(2 \log \left(\sqrt{1-4 x}+1\right)-\log (4 x)\right) \bigg)
\end{eqnarray}
This expression has interference features similar to the ones present in the case of the scalar model \citep{Bell:2016ekl}, and in particular the cross section is vanishing when the mediators are degenerate, as $\lim_{y\rightarrow x}G(x,y)=0$. The cross section is also vanishing in the limit that the pseudoscalar mixing angle goes to zero\footnote{For the Lagrangain of \ref{eq:s2hdm}, the $aah$ vertex is absent in the $\mu_{hHP}\rightarrow 0$ (and hence $\sin2\theta \rightarrow 0$) limit.  More generally, the presence in the Lagrangian of terms such as $\lambda_{P1} P^2 (\Phi_1^\dagger \Phi_1)$ or
$\lambda_{P2} P^2 (\Phi_2^\dagger \Phi_2)$ would allow the $aah$ vertex even when the mixing angle vanishes.  It would also obscure the interference effect, in the sense that the cross section would no longer vanish when $M_A=M_a$.}.
Note that the terms proportional to $z$ arise from the terms in the 2HDM potential proportional to $\hat{\lambda}_{3,4,5}$, which can be rewritten in terms of $\lambda_{1,2,3,4,5}$ and $\tan\beta$.  Therefore, the coefficients will in general depend on all these couplings and the value of $\tan\beta$. A notable exception is when $\hat{\lambda}_3+\hat{\lambda}_4+\hat{\lambda}_5=\frac{m_h^2}{v^2}=\hat{\lambda}_1=\hat{\lambda}_2$, which is a special case of our benchmark point with $\hat{\lambda}_5=0$, for which $\lambda_i=\hat{\lambda}_i$ and there is no $\tan\beta$ dependence.  Note however that if one wants to set the masses of the CP even and charged scalars to some fixed value, one is not free to set $\hat{\lambda}_{5}=0$.

Given the loop generated interaction in \ref{eq:loopps}, the nucleon operator relevant for DD is
\begin{equation}
O_1^N = \bar{\chi} \chi \bar{N} N, \label{eq:op1}
\end{equation}
resulting in an unsuppressed, spin independent, scattering cross section.
The coefficient of the nucleon operator, $c^1_N$, is related to the coefficient $c_q$ of the quark operator $\bar{\chi}\chi\bar{q}q$ by
\begin{equation}
c_N^1 = m_N\left(\sum_{q=u,d,s} \frac{c_q}{m_q} f_{T_q}^N + \frac{2}{27} f_{T_g}\sum_{q=c,b,t}\frac{c_q}{m_q}  \right),\label{eq:c1}
\end{equation}
where the quantities $f_{T_q}^N, f_{T_g}$ are  are defined in \citep{DelNobile:2013sia}.

\subsection{Inelastic Dark Matter}
\label{sec:modelI}
We consider a model with pseudo-Dirac DM \citep{DeSimone:2010tf,Davoli:2017swj} coupled to a vector boson. A pseudo-Dirac model naturally leads to an inelastic DM structure, because the dominant interactions terms will be off-diagonal in the mass eigenstates, $\chi_1$ and $\chi_2$. As such, the elastic scattering process $\chi_1N\rightarrow \chi_1 N$ will not exist at tree level. Instead, DD can only occur if the elastic process is generated at loop level, as we discuss below, or if the inelastic $\chi_1 N \rightarrow \chi_2 N$ process is kinematically accessible.

The pseudo-Dirac Lagrangian reads
\begin{eqnarray}
\mathcal{L}
&=& \bar\Psi(i\slashed{\partial}-M_D)\Psi -\frac{1}{4} F_{\mu\nu}^{V} F_{V}^{\mu\nu} +\frac{1}{4} M^2  Z_{\mu}^{'} Z^{'\mu}
+ Q_\Psi g\bar\Psi\gamma^\mu\Psi\,Z_\mu'+ Q_q g\sum_q \bar q \gamma^\mu q \,Z_\mu' \nonumber \\
&& -\frac{m_L}{2}\left(\overline{\Psi^c} P_L\Psi+\text{h.c.}\right)
-\frac{m_R}{2}\left(\overline{\Psi^c} P_R\Psi+\text{h.c.}\right)\,, 
\label{eq:linelastic}
\end{eqnarray}
where $Q_\Psi, Q_q$ are the DM and quark $U(1)$ charges. We shall set $Q_\Psi Q_q =1$ throughout, as DD constraints do not depend on the individual charges, but only on their product. Taking $m_L=m_R=\frac{1}{2}\delta m \ll m_D$, the Majorana mass eigenstates become
\begin{eqnarray}
\chi_1&=\frac{i}{\sqrt 2}\left(\Psi-\Psi^c\right)\label{eq:chi1}\,,\\
\chi_2&=\frac{1}{\sqrt 2}\left(\Psi+\Psi^c\right)\label{eq:chi2}\,.
\end{eqnarray}
Expressed in terms of the mass eigenstates, the DM Lagrangian effectively becomes
\begin{eqnarray}
\mathcal{L} \supset \frac{1}{2}\,\bar\chi_1(i\slashed{\partial}-m_1)\chi_1+\frac{1}{2}\,\bar\chi_2(i\slashed{\partial}-m_2)\chi_2\, + i Q_\Psi g \bar{\chi}_2 \gamma^\mu Z_\mu'\, \chi_1 + i Q_\Psi g \bar{\chi}_1 \gamma^\mu Z_\mu'\chi_2, \label{eq:linelasticm}
\end{eqnarray}
where $m_1=m_D-\frac{1}{2}\delta m$ and $m_2=m_D+\frac{1}{2}\delta m = m_1+ \delta m$. For more details about the model, refer to \citep{TuckerSmith:2001hy,Cui:2009xq}. Note that, once we write the Lagrangian in terms of the mass eigenstates, it is clear there can be no elastic scattering operators due to the absence of $\chi_i$-$\chi_i$ interaction terms. 

For the inelastic scattering, $\chi_1 N \rightarrow \chi_2 N$ the relevant operator at tree level is
\begin{equation}
\tilde{O}_5^N = \bar{\chi}_1\gamma^\mu \chi_2 \bar{N}\gamma_\mu N,  \label{eq:op5}
\end{equation}
with the coefficient 
\begin{equation}
c_N^5 = 3\frac{g^2}{M^2}.
\label{eq:c5}
\end{equation}
This operator is similar to the usual $O_5^N = \bar{\chi}\gamma^\mu \chi \bar{N}\gamma_\mu N$, apart for the kinematic suppression. This operator gives rise to spin-independent cross section, but the kinematic suppression, due to the inelastic nature of the scattering, allows the scattering only on the tail of the velocity integral, depending on the value of $\frac{\delta m}{\mu}$. This effectively results in an exponential suppression of the scattering rate with $\delta m$. The differential rate is given by 
\begin{eqnarray}
\frac{dR}{dE_R} &=& \frac{\rho_\chi}{m_\chi}\sum_T \frac{\xi_T}{m_T} \int_{v_{min}} d^3u u f(u) \frac{d\sigma_T}{dE_R}\left(v_{rel},E_R\right)\,,\label{eq:velocityintegral}
\end{eqnarray}
where
\begin{eqnarray}
v_{min} &=& \frac{1}{\sqrt{2m_T E_R}}\left(\frac{m_T E_R}{\mu}+\delta m\right) \ge \sqrt{\frac{2\delta m}{\mu}}=v^*\label{eq:vmin}
\end{eqnarray}
is the minimum velocity needed to up-scatter a $\chi_1$ to the heavier $\chi_2$ state. The parameter $\mu$ is the DM-nucleon reduced mass, $m_T$ is the mass of the target nucleus and $\xi_T$ are the mass fractions defined in \citep{DelNobile:2013sia}. By considering large enough $\delta m \gtrsim 10^{-6} \mu$, the velocity integral \ref{eq:velocityintegral} is on the tail of the distribution and thus the usual strong DD limits on spin-independent cross sections can be weakened or completely evaded if $v_{min}$ exceeds the galactic escape velocity, i.e. $v^*\ge v_{esc}$.

\begin{figure}[t]
    \centering
    \includegraphics{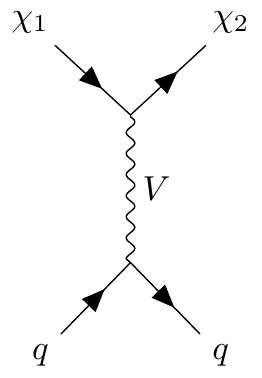} 
    \caption{Inelastic DM-nucleon scattering is kinematically suppressed unless the mass difference between $\chi_1$ and $\chi_2$ is extremely small.}
    \label{fig:feynDDtree}
\end{figure}
\begin{figure}[t]
    \centering
    \includegraphics[width=0.35\textwidth]{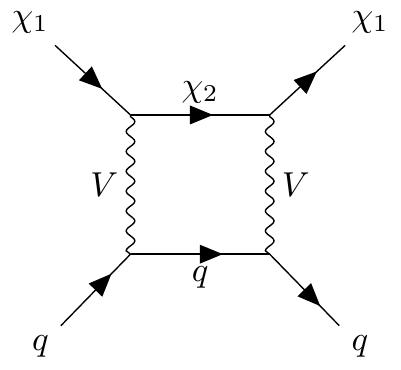} 
    \includegraphics[width=0.35\textwidth]{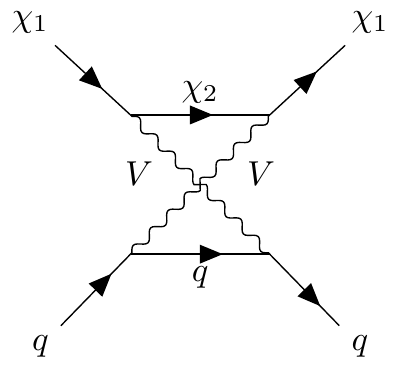} 
    \caption{Spin-independent elastic DM-nucleon scattering arises from the loop exchange of the vector mediators.}
    \label{fig:feynDDV}
\end{figure}

For mass splittings $\delta m > O(10-100 \text{keV})$, the tree level inelastic scattering process in Fig.~\ref{fig:feynDDtree} is negligible.
However, elastic scattering operators are generated at loop level, as show in Fig.~\ref{fig:feynDDV}.
The low energy effective operator is
\begin{eqnarray}
\mathcal{L}_{eff} &=&  \frac{4 g^4 m_q m_\chi}{16\pi^2 M^4}     F_3\left(\frac{m_\chi^2}{M^2}\right) \bar{\chi_1}\chi_1\bar{q}q\label{eq:loopv}\,,
\end{eqnarray}
where
\begin{eqnarray}
F_3(x) &=& \int_0^1 dz (z+1) \left(\frac{z-1}{x z^2-z+1}+\log (z (x z-1)+1)-\log (x)-2 \log (z)\right)\nonumber\\
&=& \frac{\left(8 x^2-4 x+2\right) \log \left(\frac{\sqrt{1-4 x}+1}{2 \sqrt{x}}\right)+\sqrt{1-4 x} (2 x+\log (x))}{4 \sqrt{1-4 x} x^2}\,.\label{eq:f4}
\end{eqnarray}
Because these loop diagram generate a contribution to the scalar type operator, $\overline{\chi}\chi \overline{q}q$, the resulting scattering cross section will be spin-independent. This quark operator contributes to the nucleon operator $O_1^N$ through the relation \ref{eq:c1}.

\section{Direct Detection Results}
\label{sec:constraints}

For both the pseudoscalar and inelastic model, the limits arising from tree level processes are either negligible or absent, hence we will only present the limits arising from the loop level amplitudes, with the exception of Fig.~\ref{fig:DDtreeVSloop}, where we will compare the tree and loop level cross sections. We will generate DD constraints recasting the 2016 LUX~\citep{Akerib:2016vxi} and XENON1T~\citep{Aprile:2017iyp} data, via an effective operator approach using tools from \citep{DelNobile:2013sia}.

\subsection{DD Constraints - Pseudoscalar}
\label{sec:ddconstr1}

\begin{figure}[t]
\begin{center}
\includegraphics[width=0.9\textwidth]{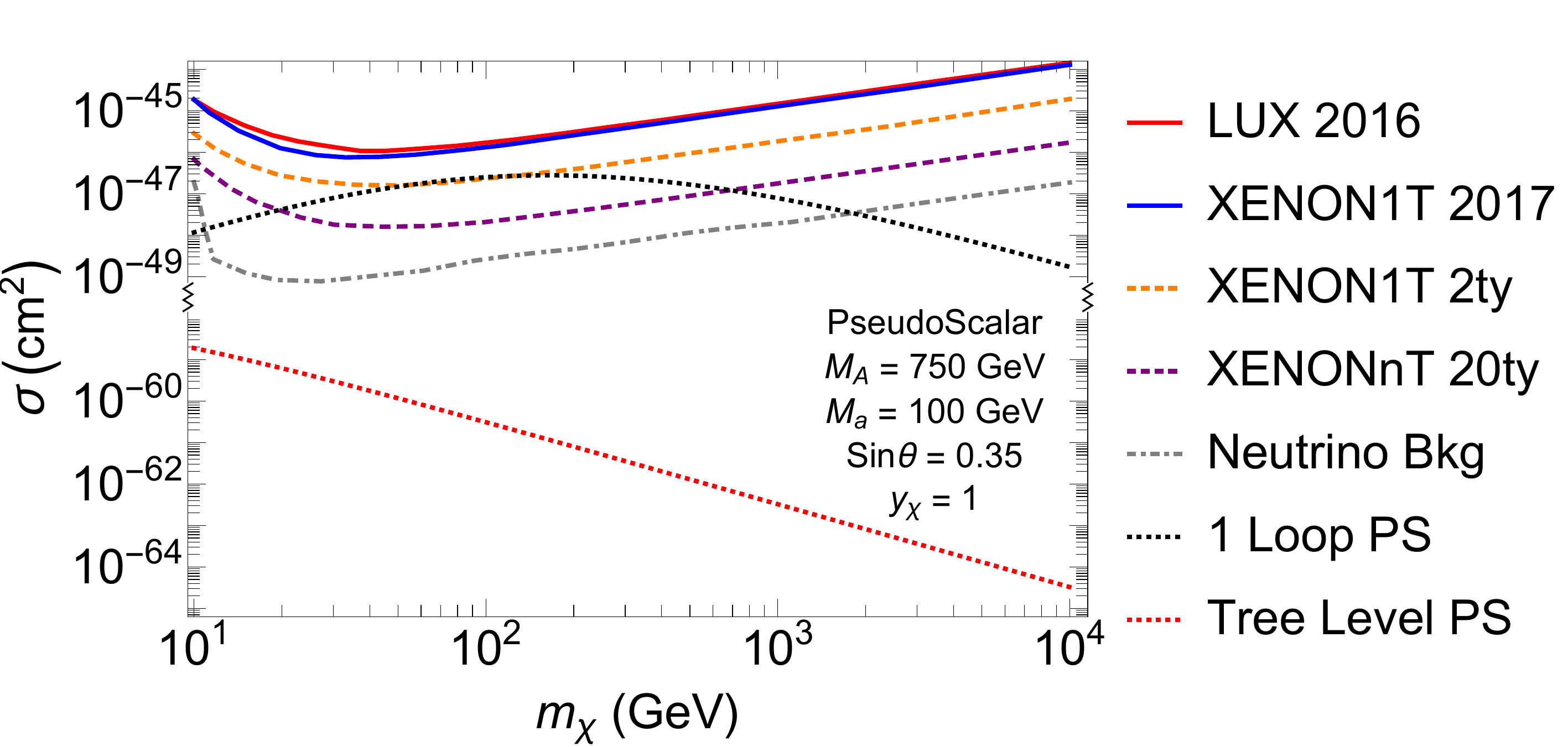}
\caption{Comparison of the loop and tree level DD cross sections for the pseudoscalar model, with current and projected experimental limits. 
The solid red and blue lines refer, respectively, to current LUX~\citep{Akerib:2016vxi} and XENON1T~\citep{Aprile:2017iyp} limits, the orange and purple dashed lines refer instead to XENON1T and XENONnT projections \citep{Aprile:2015uzo}, while the grey dot-dashed line refers to the neutrino background~\citep{Billard:2013qya}. The tree level cross section for the pseudoscalar model is shown as a red dashed line for $M_a=100\GeV$, $M_A=750\GeV$, $y_\chi=1$ and $\sin\theta=0.35$, while the loop level cross section for the same benchmark point is shown with a black dashed line.} 
\label{fig:DDtreeVSloop}
\end{center}
\end{figure}

In Fig.~\ref{fig:DDtreeVSloop} we compare the loop and tree level DD cross sections for the pseudoscalar model in the $\sigma_{p\text{-}\chi}$-$m_\chi$ plane, for a benchmark parameter point. 
Current experiments probe cross sections down to $\sigma\sim 10^{-46}\cm^2$, while projections for XENON1T and XENONnT show that these experiments should be able to probe cross sections as small as $\sigma\sim10^{-47}\cm^2$ and $\sigma\sim10^{-48}\cm^2$ respectively, for $m_\chi\sim100\GeV$. The neutrino floor (dot-dashed line) will give a background cross section $\sigma \gtrsim 10^{-49}\cm^2$. In this plot, we show the signal for a benchmark point with $M_A=750\GeV$, $M_a=100\GeV$, $\sin\theta=0.35$ and $y_\chi=1$ for the pseudoscalar model. The 1 loop level cross section is shown as a black dotted line. This point was chosen such that it lies on the boundary of the XENON1T projection, but well within the XENONnT projection, for a large range of DM masses $20\GeV < m_\chi < 700\GeV$. To stress the importance of 1-loop corrections over the tree level result, we also show the tree level scattering cross section for the same benchmark point with a red dotted line. At $m_\chi =10\GeV$, the tree level result is 11 orders of magnitude smaller than the 1 loop result. The tree level cross section also decreases approximately as $\sigma\propto m_\chi^{-2}$, resulting an even greater suppression for larger DM masses, while the one loop result obtains its dependence on $m_\chi$ from the loop functions, which make the cross section first rise, reach a maximum for $m_\chi \sim M_a$, and then fall off for larger $m_\chi$.

\begin{figure}[t]
\begin{center}
\includegraphics[width=0.49\textwidth]{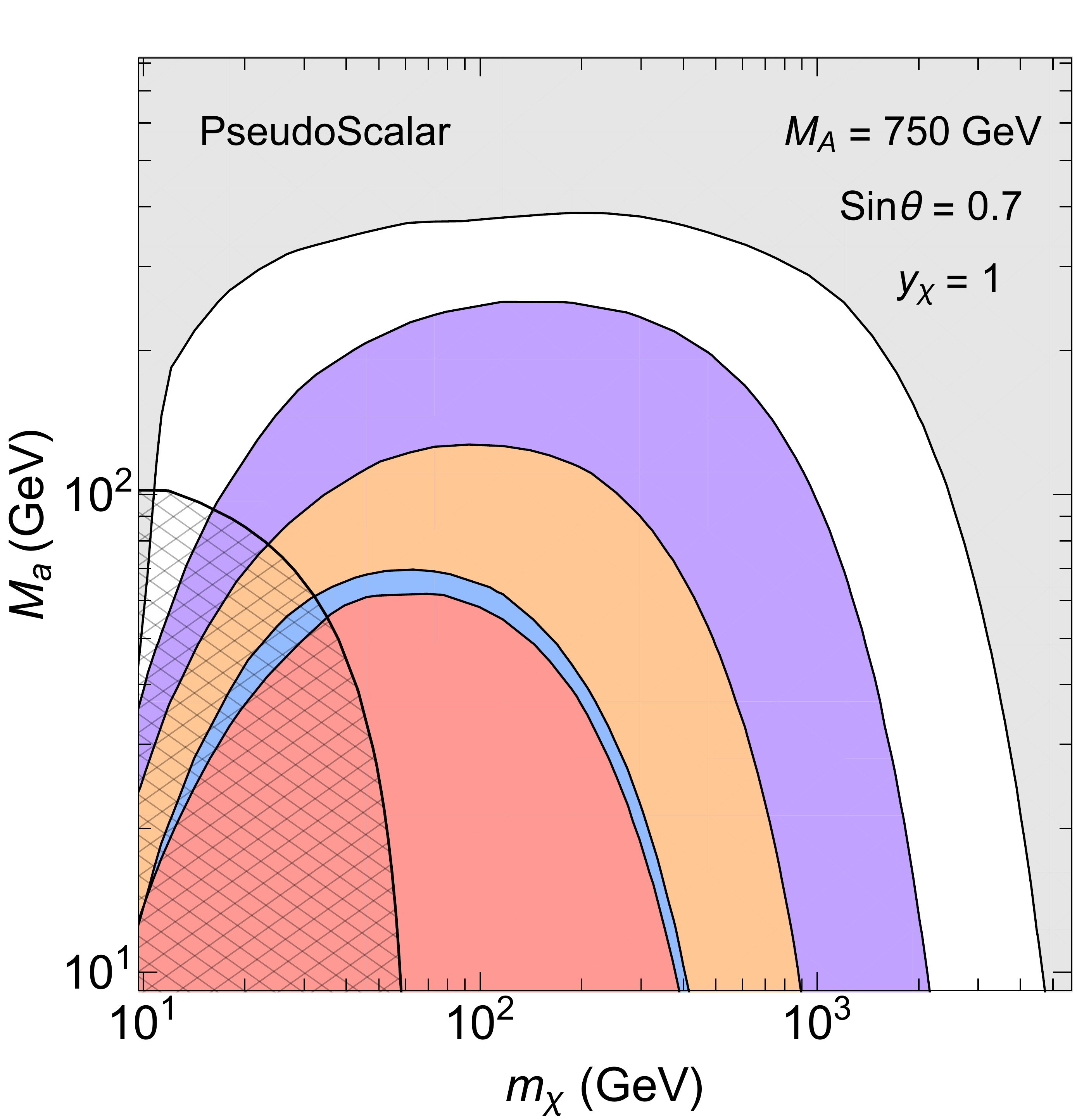}
\includegraphics[width=0.49\textwidth]{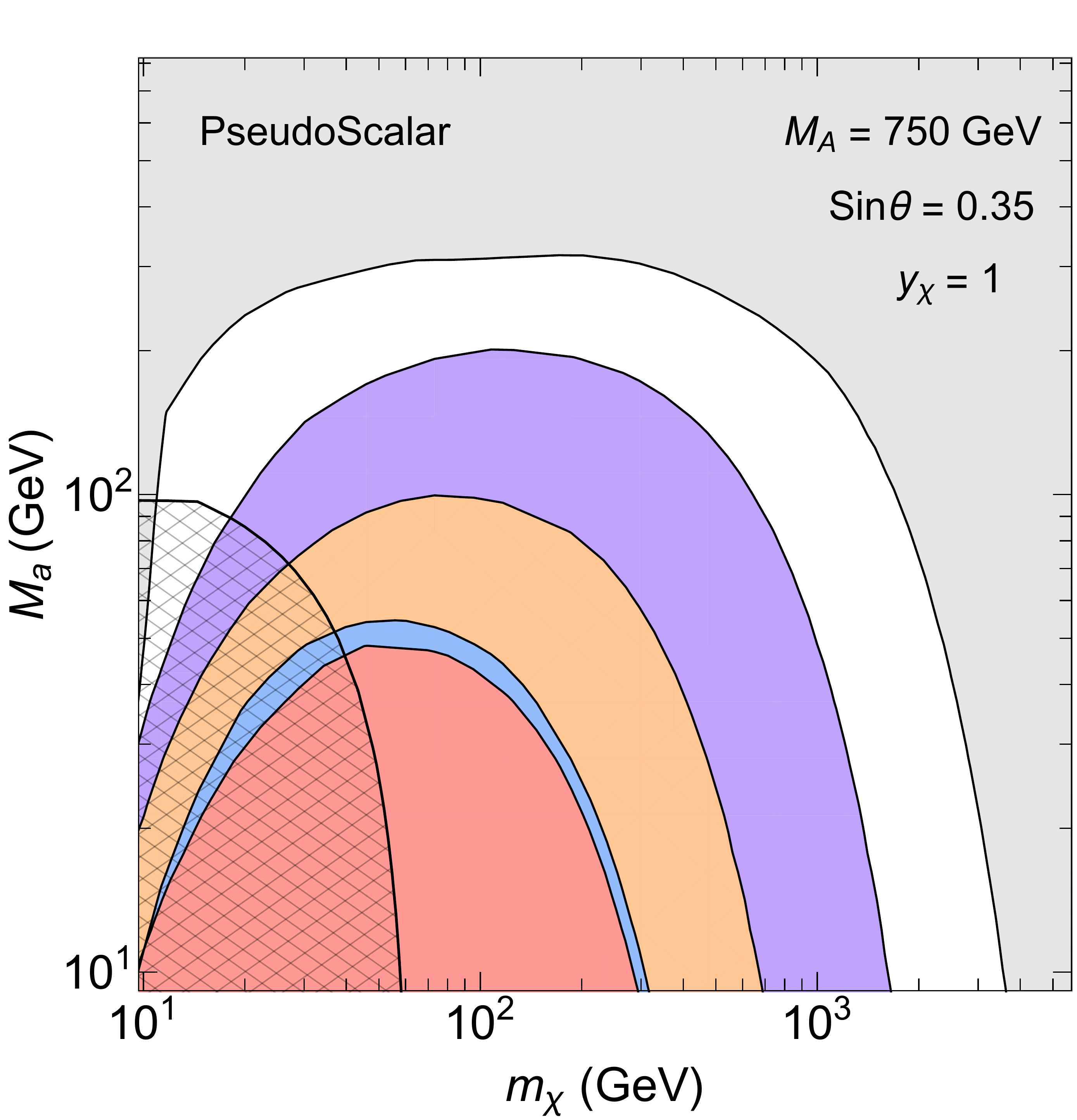}\\
\hspace{3cm}
\includegraphics[width=0.49\textwidth]{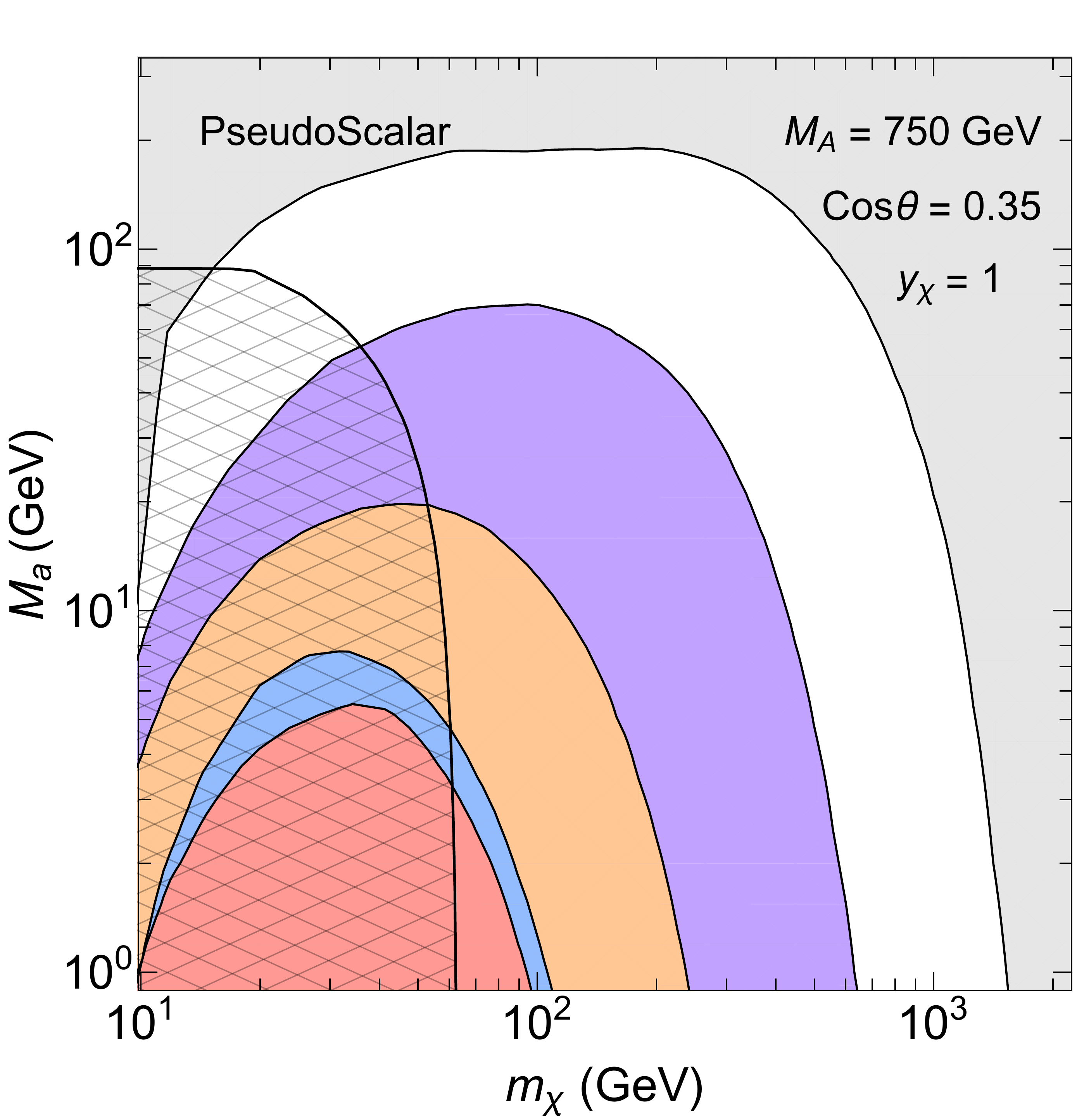}\hspace{0.5cm}
\raisebox{0.5\height}{\includegraphics[width=0.25\textwidth]{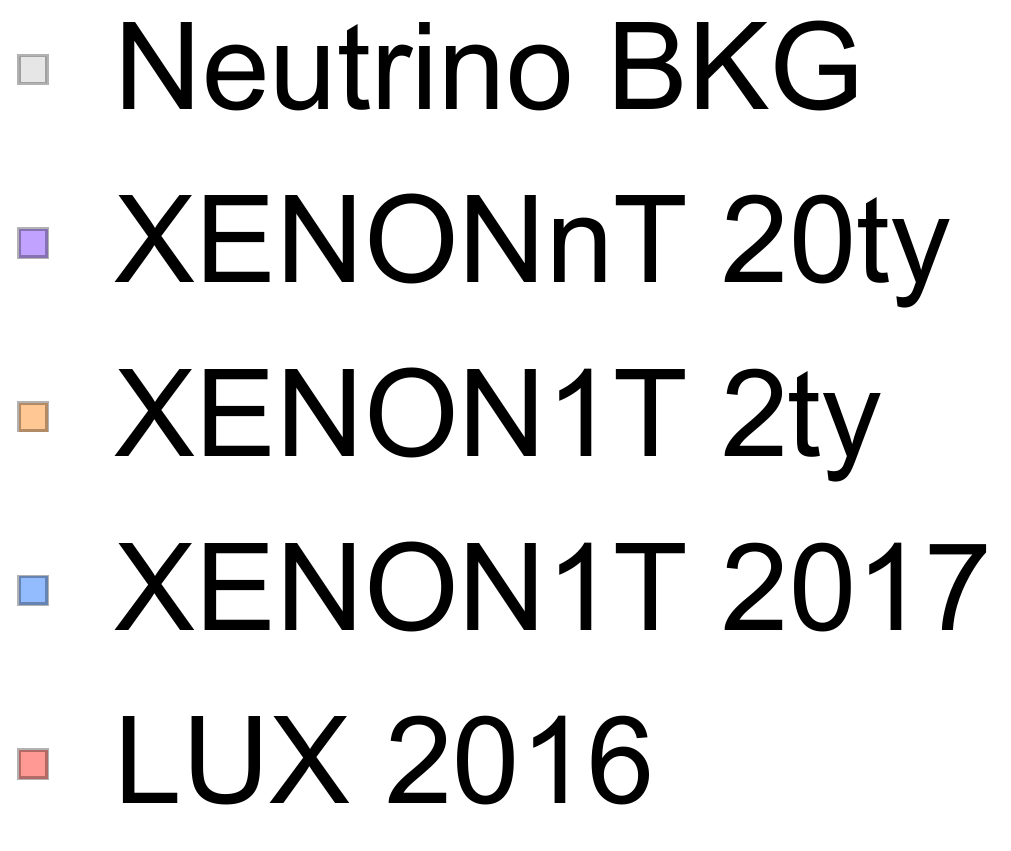}}

\caption{Direct detection exclusions and projections for the Pseudoscalar model, arising from loop contributions to the scattering amplitudes. The various regions refer, in order, to current LUX~\citep{Akerib:2016vxi} and  XENON1T~\citep{Aprile:2017iyp} limits, XENON1T and XENONnT projections~\citep{Aprile:2015uzo}, and the neutrino background \citep{Billard:2013qya}. The top left panel takes $\sin\theta=0.7$, the top right panel $\sin\theta=0.35$, and the bottom panel $\cos\theta=0.35$, while $M_A=750\GeV$ in all the panels. The hatched region is eliminated by Higgs invisible width constraints.} 
\label{fig:PSDD}
\end{center}
\end{figure} 

The current and projected DD constraints on the pseudoscalar model are presented in Fig.~\ref{fig:PSDD}. The limits were calculated using the full expression in \ref{eq:loopps}. The heavier pseudoscalar has been set to $M_A=750\GeV$, and the mixing angles fixed to $\sin\theta=0.7$ in the left panel, $\sin\theta=0.35$ in the right panel and $\cos\theta=0.35$ in the bottom panel. The different colors refer to different experiments: red corresponds to LUX \citep{Akerib:2016vxi}, light blue to XENON1T\citep{Aprile:2017iyp}, yellow and purple are projections for XENON1T and XENONnT respectively \citep{Aprile:2015uzo}, and the grey shaded area is the one not accessible to ordinary DD experiments due to the presence of the Neutrino background\citep{Billard:2013qya}. The hatched regions are excluded by limits from the Higgs invisible decay width~\citep{Aad:2015pla,Khachatryan:2016whc}, arising from 2 and 3 body decays as described in \citep{Bauer:2017ota}. They rule out the low $m_\chi,M_a$ mass region. In the top left panel ($\sin\theta=0.7$), current limits are able to exclude the portion of parameter space with $10\GeV \lesssim m_\chi \lesssim 400\GeV$ and $M_a \lesssim 60\GeV$. Projected limits for XENON1T and XENONnT could expand the excluded region to $10\GeV \lesssim m_\chi \lesssim 2\TeV$ and $M_a \lesssim 200\GeV$. The presence of the neutrino floor will prevent to be able to probe this model for $M_a\gtrsim 350\GeV$ or $m_\chi \gtrsim 4\TeV$ with conventional DD experiments. The top right panel ($\sin\theta=0.35$) is quite similar to the first, with the limits just slightly weakened by the smaller mixing angle. In the lower panel ($\cos\theta=0.35$) current DD experiments can only probe a tiny portion of the parameter space, with $10\GeV \lesssim m_\chi \lesssim 100\GeV$ and $M_a \lesssim 6\GeV$. In this case we see that, indeed, most of the region currently ruled out by DD is also eliminated by Higgs width constraints. Projected limits expand the range to up $M_a\sim 60\GeV$ and $m_\chi\sim 600\GeV$, while neutrino background makes inaccessible to DD the region beyond $M_a\gtrsim 200\GeV$ or $m_\chi \gtrsim 1.5\TeV$. 

\begin{figure}[t]
\begin{center}
\includegraphics[width=0.49\textwidth]{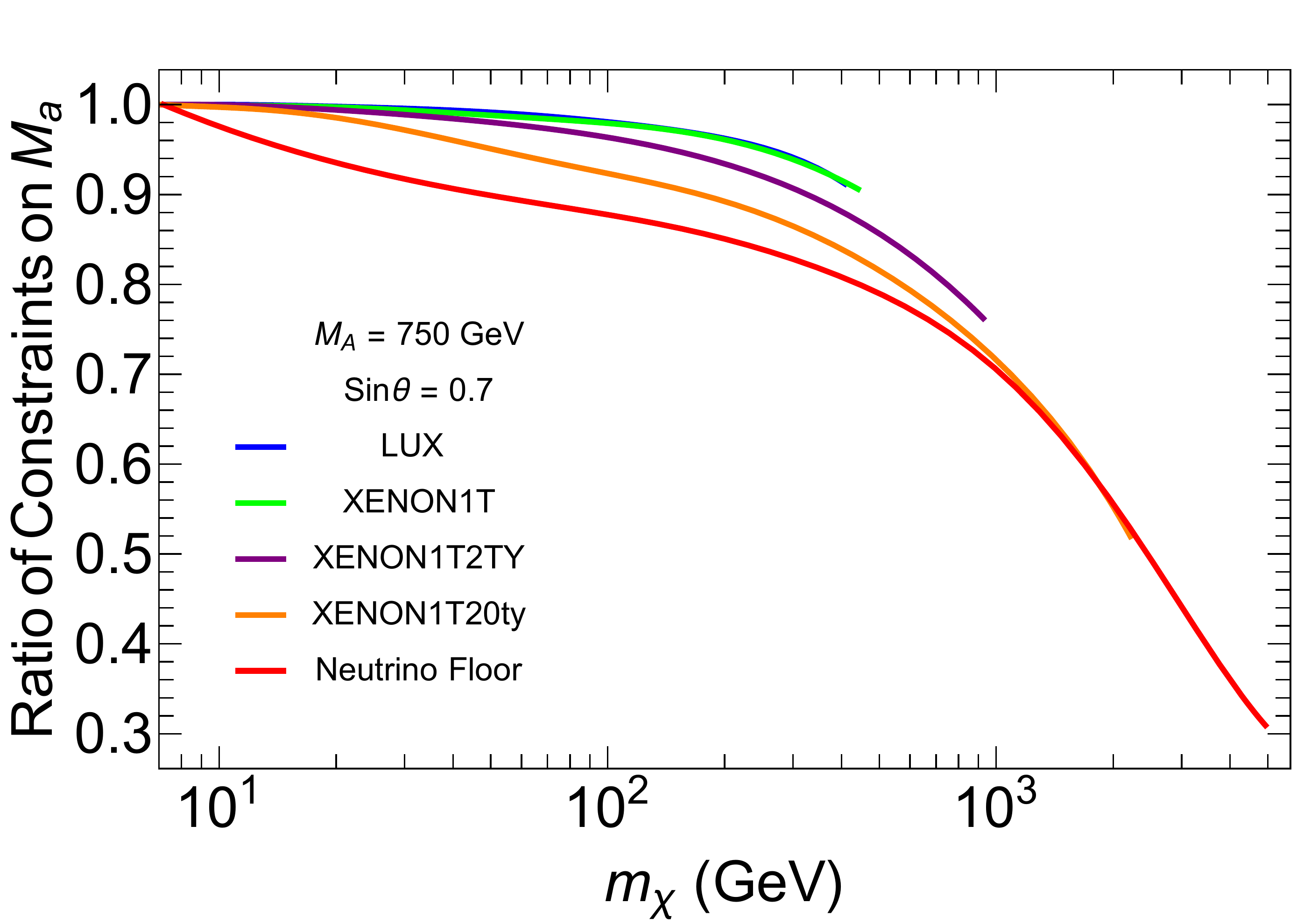}
\includegraphics[width=0.49\textwidth]{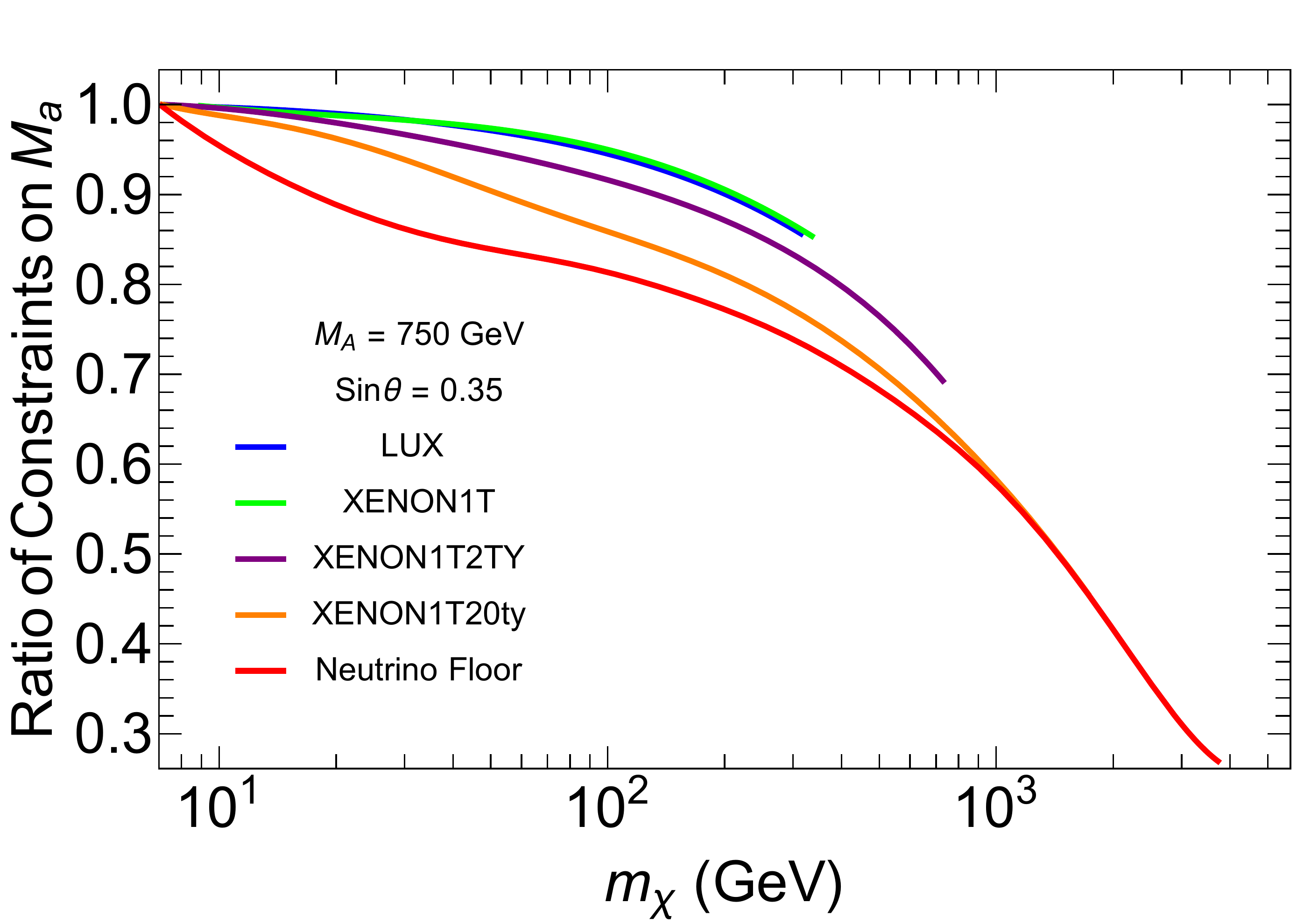}\\
\includegraphics[width=0.49\textwidth]{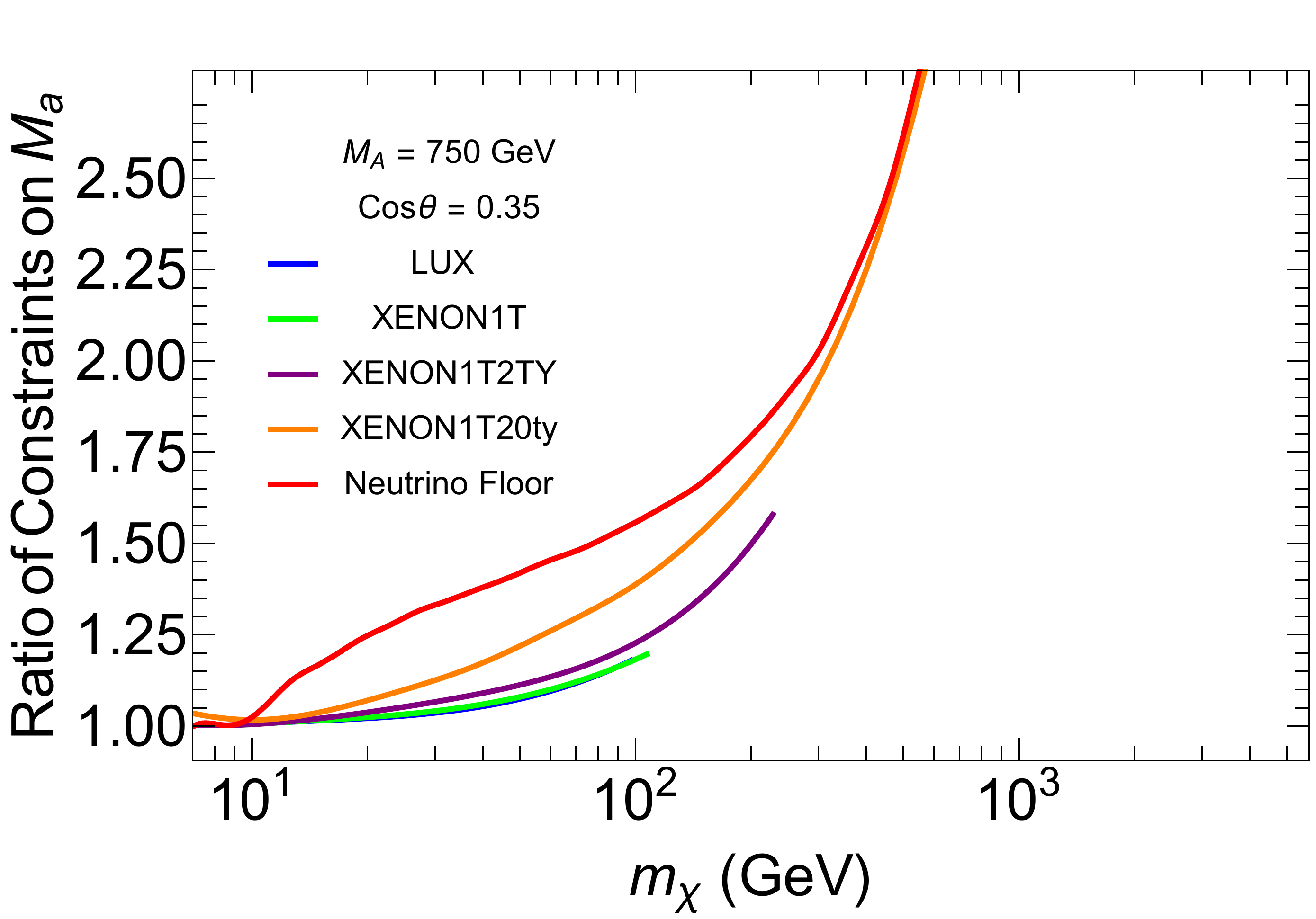}
\caption{Accuracy of the full vs approximated amplitude for the pseudoscalar model.  The curves show the ratio of $M_a$ limits or projections using the full to approximated expressions of \ref{eq:loopps} and \ref{eq:looppsapprox}, respectively. The various curves refer, in order, to LUX~\citep{Akerib:2016vxi} (blue) and XENON1T~\citep{Aprile:2017iyp} (green) limits, XENON1T (purple) and XENONnT (orange) projections~\citep{Aprile:2015uzo}, and the neutrino background~\citep{Billard:2013qya} (red). The top left panel takes $\sin\theta=0.7$, the top right panel $\sin\theta=0.35$, and bottom panel $\cos\theta=0.35$, while $M_A=750\GeV$ is used in all the panels.} 
\label{fig:PScomp}
\end{center}
\end{figure} 

We also investigate the accuracy of using the approximation \ref{eq:looppsapprox} rather than the full expression of \ref{eq:loopps}. The ratio of the bounds on $m_a$ obtained using the full expression, to the ones using the approximated expression, are shown in Fig.~\ref{fig:PScomp}. One can see that using the approximated expression \ref{eq:looppsapprox} always gives stronger constraints than the full expression when the mixing angle is small. Conversely, the situation is reversed when the mixing angle is large. This is principally because of two reasons: Firstly, the full expression includes the interference between the 2 propagators, while the approximated one does not. Interference will mostly be important when the masses of the two mediators are comparable, so if we set $M_A=750\GeV$, interference will start to be important when $M_a$ is a few hundred $\GeV$. Thus the approximation will always be worse for stronger limits, that are able to probe values of $M_a$ closer to $M_A$ (we remind the reader that the DD cross section goes to zero for $\lim_{M_a\rightarrow M_A}$, and this feature can be easily noted in Fig.~\ref{fig:PScomp}). Secondly, if the mixing angle is large and close to $\pi/2$, then the diagrams containing the heavier scalar become more important than the ones with only the light scalar (of which the diagram is suppressed by $\cos^2 \theta$), and can actually dominate the result in some regions of the parameter space, especially at large $m_\chi$. As one can see from Fig.~\ref{fig:PScomp}, the heavier the DM is, the worse the approximation becomes.

\begin{figure}[t]
\begin{center}
\includegraphics[width=0.49\textwidth]{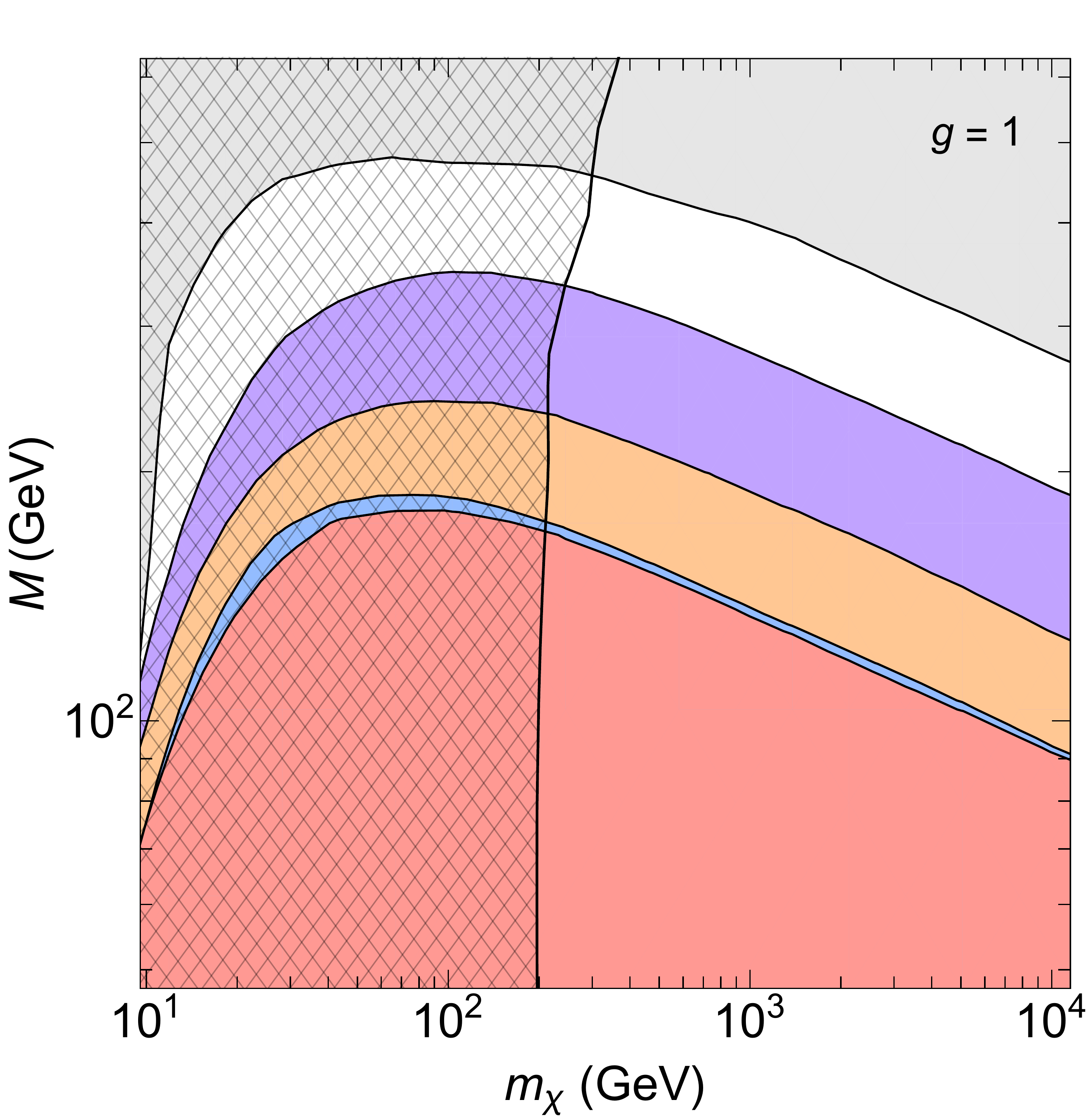}
\includegraphics[width=0.49\textwidth]{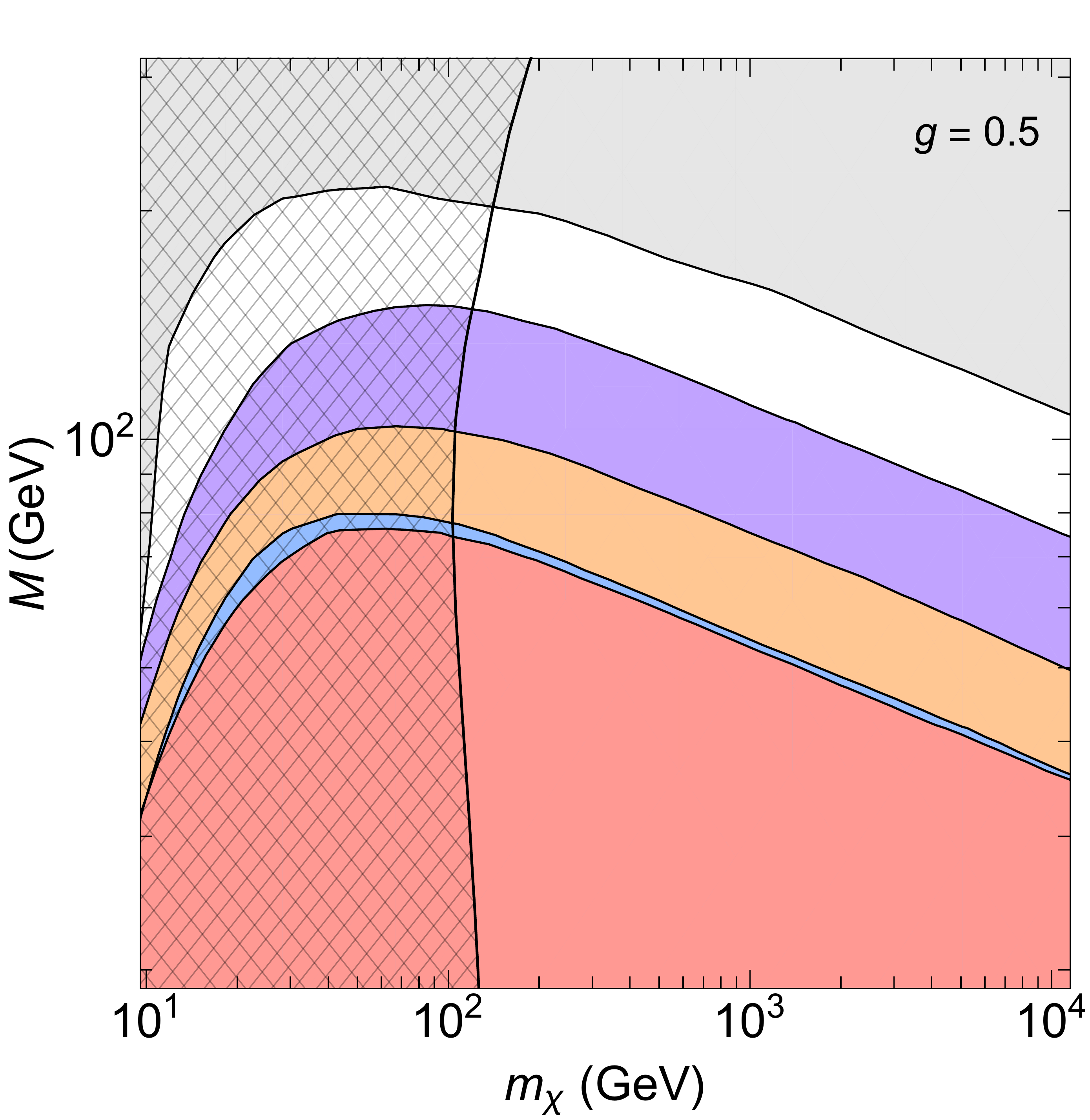}\\
\hspace{3cm}
\includegraphics[width=0.49\textwidth]{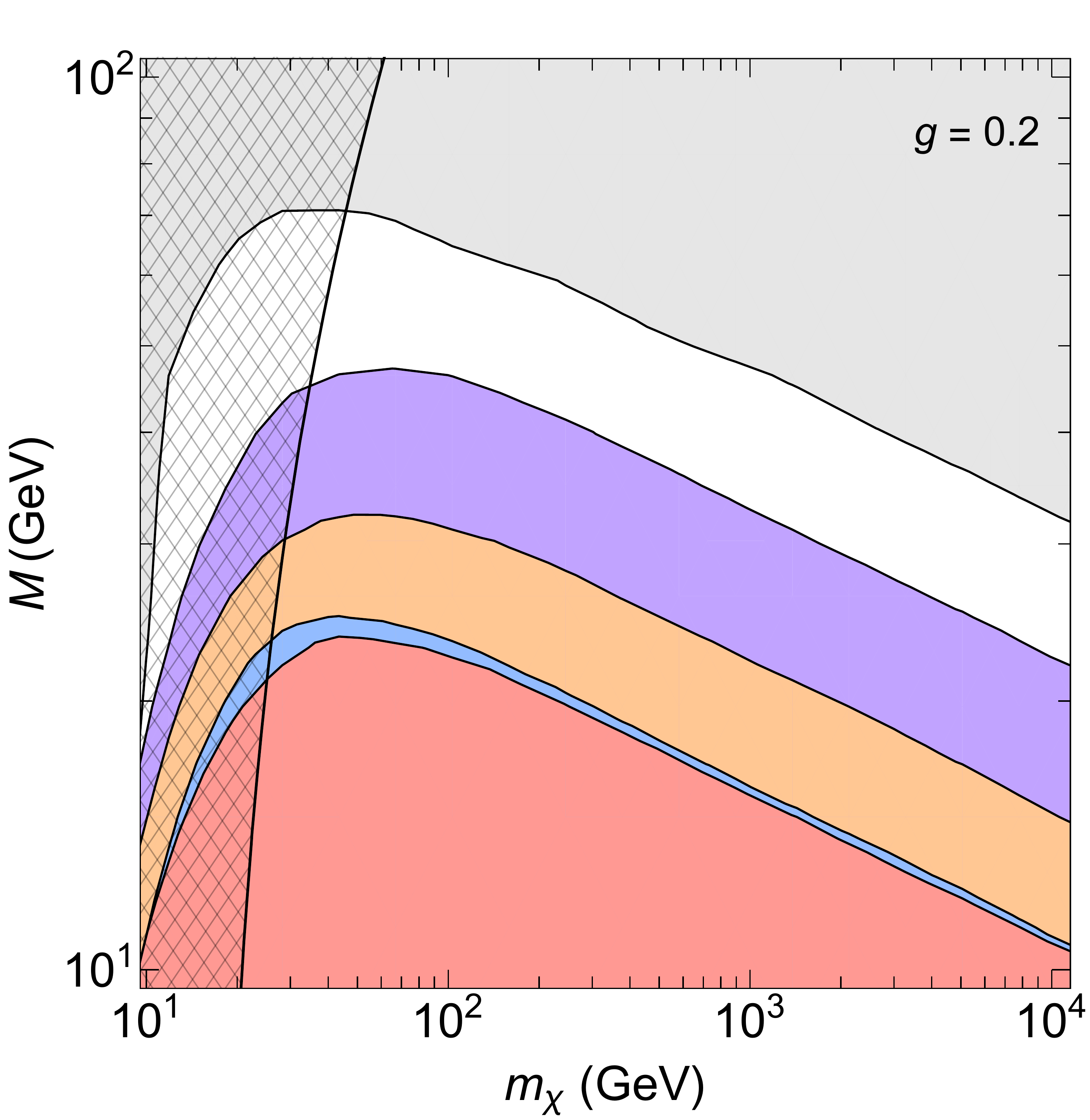}\hspace{0.5cm}
\raisebox{0.5\height}{\includegraphics[width=0.25\textwidth]{plots/legend.pdf}}
\caption{Direct detection exclusions and projections for the Inelastic model, arising from loop contributions to the elastic scattering rate. The various regions refer, in order, to current LUX`\citep{Akerib:2016vxi} and XENON1T~\citep{Aprile:2017iyp} limits, XENON1T and XENONnT projections~\citep{Aprile:2015uzo}, and the neutrino background \citep{Billard:2013qya}. The top left panel takes $g=1$, the top right panel $g=0.5$, and the bottom panel $g=0.2$. The hatched region indicates the limits from LHC monojet searches, using results from ATLAS~\citep{Aaboud:2017phn}. To determine the excluded regions for the upper left and the bottom panels, we rescaled the ATLAS~\citep{Aaboud:2017phn} limits.} 
\label{fig:INDD}
\end{center}
\end{figure} 

\subsection{DD Constraints - Inelastic DM}
\label{sec:ddconstr2}

Direct detection constraints for the inelastic model are presented in Fig.~\ref{fig:INDD}. The coupling constant $g$ is set to $g=1$ in the left panel, $g=0.5$ in the right panel and $g=0.2$ in the bottom panel. In the first case, current experiments exclude values of $M$ below $100\GeV$ to $200\GeV$, for $10\GeV<m_\chi<10\TeV$. Projected limits will be able to reach values of $M$ up to $400\GeV$ to $500\GeV$ in the same DM mass range. The neutrino background will prevent values of $M$ larger than $600\GeV$ to $800\GeV$ to be probed with conventional DD experiments. In the second case, current experiment exclude values of $M$ below $50\GeV$ to $90\GeV$, for $10\GeV<m_\chi<4\TeV$. Projected limits will be able to reach values of $M$ up to $200\GeV$ in the same DM mass range, while the neutrino background will prevent us from exploring values of $M$ larger than $200\GeV$ to $400\GeV$, for $10\GeV<m_\chi<10\TeV$. In the third case, limits from current experiments are pushed down to $M\sim20$-$30\GeV$, projected limits to $M\sim50$-$70\GeV$ and the neutrino floor to $M\sim 70$-$150\GeV$. In the 3 panels we also show limits arising from LHC monojet searches, using the results from ATLAS~\citep{Aaboud:2017phn}. To calculate monojet limits for the upper left panel and the bottom panel, we rescaled the limit from ATLAS~\citep{Aaboud:2017phn} by implementing the model in Madgraph~\citep{Alwall:2014hca} using Feynrules~\citep{Christensen:2008py,Alloul:2013bka}, to calculate the monojet cross section. In doing this, we assume that the mass splitting $\delta m$ is not large enough to produce displaced vertices. As Monojet constraints also depend on the individual charges of $\chi$ and $q$, we assume $Q_\Psi/Q_q=4$, as done by CMS and ATLAS collaborations \citep{Boveia:2016mrp,Albert:2017onk}.

\section{Conclusions}
\label{sec:conclusions}


We have analyzed effects coming from EFT operators that arise at loop level once the usual tree level operators are embedded in a full gauge invariant and renormalisable theory, and we have done this in the context of two example models. 
The first model is the minimal scenario featuring a pseudoscalar mediator connecting the dark and visible sectors in a gauge-invariant way, which necessarily involves \emph{two} pseudoscalar mediators. 
The second model is that of inelastic DM coupled to the SM through a spin-1 mediator with purely vector couplings. Both models are subject to very weak DD constraints when considering only EFT operators generated at tree level: the former generates a spin-dependent cross section, suppressed by the fourth power of the transferred momentum $q_{tr}^4$, while the latter can completely avoid DD constraints if the mass splitting is large enough to kinematically suppress inelastic scattering. By calculating the operators induced at one loop in both models, we have demonstrated that a spin-independent elastic cross section is induced. 
The loop suppression factor is small compared with the suppression of the tree level contribution, and thus loops dominate the scattering cross section. 

In the case of inelastic DM, the EFT operator generated is the four fermion interaction $\bar{\chi}\chi\bar{q}q$, while in the case of the pseudoscalar model one generates an effective interaction between the SM Higgs and the DM, $h\bar{\chi}\chi$. Because of this, DD rates do not depend on the choice of the Yukawa sector for the 2HDM, making this a powerful way to probe scenarios where the couplings of the additional doublet are small or vanishing, as in the Type I 2HDM with large $\tan\beta$, or in the inert 2HDM.

We have calculated current and projected DD limits arising from these loop level scattering amplitudes. Despite being loop suppressed, we are able to probe parameter space that was previously considered inaccessible to DD.
For the pseudoscalar model, we can currently exclude mediator masses $M_a \lesssim \mathcal{O}(50 \GeV )$ and DM mass of $\mathcal{O}(20 \GeV) \lesssim m_\chi\lesssim \mathcal{O}(400 \GeV )$, for large mixing angles. 
Future DD experiments, however, can reach $M_a$ values of a few hundred GeV and DM masses up to $\mathcal{O}(\TeV)$. 
In the case of the Inelastic DM model, for a coupling of $g=1$, experimental data from DD, together with mono-jet limits, are able to exclude mediators with $M\lesssim \mathcal{O}(100\GeV)$ for DM masses $m_\chi < 10 \TeV$; projected DD limits could strengthen this by a factor of 2.

Previous analyses of the pseudoscalar scenario neglected all diagrams bar the one containing only the lightest pseudoscalar mediator. In fact, the diagrams containing the heavier pseudoscalar can be important, particularly when $m_\chi$ is large. If the mixing angle is large, the single mediator approximation overestimates the constraints. Conversely, if the mixing angle is small, the true constraints are stronger than those derived in the single mediator approximation.



\section*{Acknowledgements}
NFB and GB were supported in part by the Australian Research Council, and and IWS by the Commonwealth of Australia. 
Feynman diagrams were drawn using TikZ-Feynman\citep{Ellis:2016jkw}. We acknowledge helpful correspondence with David McKeen, Giorgio Arcadi, Manfred Lindner, Farinaldo S. Queiroz, Werner Rodejohann and Stefan Vogl. 

\newpage
\appendix

\section{Loop Functions}
\label{sec:loops}

\begin{eqnarray}
F_1\left(\hat{x}\right) &=& 16\pi^2 i \int \frac{d^4k}{(2\pi)^4} \frac{-k^\mu e_{1,\mu}}{\left((e_1+k)^2-1\right)\left(k^2-\frac{1}{\hat{x}}\right)^2}\\
F_2\left(\hat{x},\hat{y}\right) &=& 16\pi^2 i \int \frac{d^4k}{(2\pi)^4} \frac{-k^\mu e_{1,\mu}}{\left((e_1+k)^2-1\right)\left(k^2-\frac{1}{\hat{x}}\right)\left(k^2-\frac{1}{\hat{y}}\right)}
\end{eqnarray}
where $e_1^\mu=(1,0,0,0)$ and $k$ is dimensionless. They can be simplified to
\begin{eqnarray}
F_1\left(\hat{x}\right) &=& \hat{x}\int_0^1 dx\int_0^{1-x}dy \frac{x}{\hat{x}x^2+(1-x)} =  \hat{x}\int_0^1 dx \frac{x(1-x)}{x^2\hat{x}+(1-x)}\\
F_2\left(\hat{x},\hat{y}\right) &=& \hat{x}\hat{y}\int_0^1 dx\int_0^{1-x}dy \frac{x}{x^2 \hat{x}\hat{y}+y \hat{y} + (1-x-y) \hat{x}}\nonumber\\
&=&  \frac{\hat{x}\hat{y}}{\hat{y}-\hat{x}} \int_0^1 dx x\log\frac{(1-x)\hat{y}+x^2 \hat{x}\hat{y}}{(1-x)\hat{x}+x^2 \hat{x}\hat{y}}.
\end{eqnarray}
$G$ can be rewritten as
\begin{eqnarray}
G\left(x,y,z,\theta\right) &=& \frac{1}{y}A_1(x)\sin^2\theta (2xyz\cos^2\theta+(y-x)\sin^2 2\theta) \nonumber\\
&+&  \frac{1}{x}A_1(y)\cos^2\theta (2xyz\sin^2\theta+(x-y)\sin^2 2\theta) \nonumber\\
&+& \frac{\sin^2 2\theta}{x-y}\left(xyz+(x-y)\cos2\theta\right)\left(B_1(x)-B_1(y)\right)\label{eq:loopfull}
\end{eqnarray}
with
\begin{eqnarray}
A_1(x) &=&\frac{-2 \left(\sqrt{1-4 x}+1\right) (3 x-1) \log \left(\frac{\sqrt{1-4 x}+1}{2 \sqrt{x}}\right)-\left(-4 x+\sqrt{1-4
   x}+1\right) ((x-1) \log (x)-2 x)}{ 4\left(-4 x+\sqrt{1-4 x}+1\right) x^2}\qquad\enskip\\
B_1(x) &=&\frac{2 \sqrt{1-4 x} \log \left(\sqrt{1-4 x}+1\right)+2 x (1-\log (x))+\log (x)-\sqrt{1-4 x} \log (4 x)}{8 x^2}.
\end{eqnarray}
We calculate the 2 box diagrams in the zero momentum approximation for the quark\footnote{For more details, please check \citep{Cirelli:2005uq}.}. This corresponds to neglecting higher-derivative operators. We also set the quark mass to zero in the denominator, thus neglecting terms of $\mathcal{O}(m_q^2)$. By simplifying the 2 fermion lines and contracting the indices, we get:
\begin{eqnarray}
8m_\chi m_q \mathbb{1}_{ij} \mathbb{1}_{kl} -4m_q \gamma^\mu_{ij} (k+p_\chi)_\mu \mathbb{1}_{kl}+4im_q \epsilon_{\mu\nu\rho\sigma}\gamma^\mu_{ij} (k+p_\chi)_\sigma \sigma^{\nu\rho}_{kl}+8m_\chi m_q \sigma^{\mu\nu}_{ij} \sigma_{\mu\nu, kl}.
\end{eqnarray}
The last 2 terms will generate momentum-suppressed operators, and thus we discard them. The first two terms will instead generate the usual SI operator $\bar{\chi}\chi\bar{q}q$. To find the coefficient we thus calculate
\begin{eqnarray}
F_3\left(\hat{x}\right) &=& 16\pi^2 i \int \frac{d^4k}{(2\pi)^4} \frac{1-k^\mu e_{1,\mu}}{\left((e_1+k)^2-\hat{x}\right)\left(k^2\right)\left(k^2-1\right)^2}= \int_0^1 dx\int_0^{1-x}dy \frac{\left(1+x\right)\left(1-x-y\right)}{\left(x^2\hat{x}+1-x-y\right)^2}\nonumber\\
&=& \int_0^1 dx \left(1+x\right)\left(\frac{x-1}{x^2\hat{x}-x+1}-\log \hat{x} -2\log x+\log\left(1-x+x^2\hat{x}\right)\right).
\end{eqnarray}


\newpage

\label{Bibliography}

\lhead{\emph{Bibliography}} 

\bibliography{Biblio} 

\end{document}